\documentclass[journal]{IEEEtran}
\usepackage{amsmath,amsfonts,amssymb}
\usepackage{graphicx}
\usepackage{cite}
\usepackage{algorithm}
\usepackage{algorithmic}
\usepackage{booktabs}
\usepackage{multirow}
\usepackage{array}
\usepackage[colorlinks,citecolor=blue,linkcolor=blue]{hyperref}
\graphicspath{{figures/}}
\newcommand{\C}{\mathbb C}
\newcommand{\R}{\mathbb R}
\newcommand{\diag}{\operatorname{diag}}
\newcommand{\supp}{\operatorname{supp}}
\newcommand{\soft}{\mathcal S}
\newcommand{\tr}{\operatorname{tr}}
\newcommand{\vect}{\operatorname{vec}}
\newcommand{\blkdiag}{\operatorname{blkdiag}}

\newtheorem{proposition}{Proposition}
\newtheorem{corollary}{Corollary}
\newtheorem{remark}{Remark}

\begin{document}

\title{Electromagnetic-Twin Channel Estimation: From Sparse Pilots to Persistent CSI}

\author{Tuo~Wu, 
        K. C. Ho,~\IEEEmembership{Fellow,~IEEE} %
\thanks{Corresponding author: K. C. Ho.}
\thanks{T. Wu is with the School of Electronic and Information Engineering, South China University of Technology, Guangzhou 510640, China (E-mail: $\rm  wutuo@scut.edu.cn$). 	K. C. Ho is with the Department of Electrical Engineering and Computer Science, University of Missouri, Columbia, MO 65211, USA (E-mail: $\rm 	dkcho@ieee.org$).}}

\markboth{}{Wu \MakeLowercase{\textit{et al.}}: Persistent Electromagnetic-Twin Channel Learning}
\maketitle

\begin{abstract}
Each repeated pilot block is typically treated as a separate channel-estimation
problem, although much of the angle--delay state persists across updates.
This paper defines an electromagnetic twin (ET) as a recursively synchronized
signal-processing state containing channel state information (CSI),
coefficientwise uncertainty, support confidence, and age. A channel knowledge
map may initialize this state; subsequent pilots predict, correct, validate,
and store the state. We formulate an anchored-window inverse problem that yields
online uncertainty-weighted correction and fixed-lag smoothing as two convex
operating modes of a persistent-state estimator. Uncertainty-aware ET FISTA
(UET-FISTA) gives a zero-delay update, while an alternating direction method
of multipliers (ET-ADMM) jointly refines a short window when delayed CSI is
acceptable. Held-out pilots choose update, complete refresh, or hold under a
prescribed family-wise false-acceptance probability, preventing an inaccurate
stored state from being reused blindly. Conditional Bayesian Cram\'er--Rao
bounds quantify the information supplied by the stored state and by future
pilot blocks. With 20\% pilots, UET-FISTA reaches $-11.55$ dB normalized
mean-square error, compared with $-3.42$ dB for static FISTA. Over 200
independent 50-epoch trajectories, three-epoch smoothing improves matched-state
NMSE from $-10.28$ to $-13.41$ dB. On 40 off-grid QuaDRiGa trajectories,
online UET-FISTA improves static recovery from $-1.37$ to $-1.88$ dB and
effective rate from 1.688 to 1.919 bit/s/Hz. These results identify when
persistent CSI reduces pilot demand and when the receiver should instead
refresh or preserve its state.
\end{abstract}


\section{Introduction}
\IEEEPARstart{A}{ccurate} channel state information (CSI) supports coherent
detection, precoding, interference suppression, localization, and sensing in
wideband and large-array wireless systems. These functions become more
dependent on timely CSI as emerging 6G designs introduce larger apertures,
wider bandwidths, and reconfigurable antenna or propagation
states~\cite{wuFluid}. Fortunately, many physical channels admit compact
representations in angle--delay or angle--delay--Doppler domains. Compressed
sensing can therefore recover useful CSI from fewer pilots than conventional
least-squares (LS) estimation
~\cite{donohoCS,bajwaCCS,heathMmWave,alkhateebMmWave}. Related work in
\emph{IEEE Transactions on Signal Processing} has further developed sparse
multicarrier recovery, low-dimensional channel-subspace estimation, and
compressive estimation for large and wideband arrays
~\cite{bergerSparse,haghighatshoarSubspace,tsaiCompressive,wangBeamSquint}.
Most sparse estimators, however, are implemented as snapshot methods: one pilot
block enters the receiver, one channel estimate is produced, and the next
block starts another recovery problem.

That snapshot view is safe but forgetful. Consider a receiver that has already
observed the same link over several sounding epochs. A dominant reflector may
remain at nearly the same angle and delay, while its complex gain changes
gradually. A moving person may temporarily block one path, and opening a door
may reveal another. Channel aging and angle--delay prediction studies confirm
that such temporal evolution directly affects the usefulness of stored
CSI~\cite{truongAging,wuPrediction}. Estimating the full channel from zero at every epoch
discards the persistent part of this history. Reusing the last estimate
without verification makes the opposite mistake: it saves pilots when the
environment is stable but can propagate a stale path or a biased gain after
an abrupt event. Practical operation lies between these two extremes. The
receiver should reuse what remains credible, spend the newest pilots on what
has changed, and recognize when the stored information should no longer be
trusted.

This observation broadens the scope of channel estimation. The task is no
longer only to recover the current channel from the current pilots. It is to
maintain a wireless state across time. At each epoch, the receiver begins
with a predicted channel, compares it with new measurements, estimates their
mismatch, validates the resulting candidate, and decides what information
will be available at the next epoch. The stored object should include more
than a point estimate. It should also record which coefficients are
uncertain, which propagation components have remained active, and how long
each item has gone without convincing measurement support. These quantities
determine whether memory reduces the current estimation burden or instead
creates negative transfer.

Reusable propagation information also appears in channel knowledge maps
(CKMs) and wireless digital twins. A CKM associates a location or environment
descriptor with channel knowledge and can provide a useful spatial prior
before a link is sounded~\cite{zengCKM,zengCKMTutorial}. Channel charting
similarly extracts a geometry-preserving representation from accumulated
CSI, illustrating another way in which radio measurements can become reusable
environment knowledge~\cite{studerChart}. A wireless digital
twin has a broader system role: it may combine geometry, devices, traffic,
resources, and control variables to support network monitoring and
decision-making~\cite{alkhateebTwin,wuDTSurvey,khanDT6G}. The object studied here is narrower and
operates at the signal-processing timescale. We define an
\emph{electromagnetic twin} (ET) as a persistent, measurement-synchronized
electromagnetic state that is updated by pilots and queried by a communication
function. In simple terms, a CKM can tell the receiver what channel to expect,
the ET records what the link currently knows and how reliable that knowledge
is, and a broader digital twin can use the ET together with non-channel
network states. The three concepts are therefore complementary rather than
competing names for the same object.

Sequential channel recovery is, of course, not new. Kalman-filtered compressed
sensing, dynamic approximate message passing, and temporally correlated sparse
Bayesian learning (SBL) exploit temporal signal
statistics~\cite{vaswaniKFCS,zinielDCS,zhangTSBL}. Prior-support and
time-weighted sparse estimators provide related convex mechanisms for carrying
information across inverse problems~\cite{vaswaniModifiedCS,angelosanteRLS},
while learned channel estimators can infer spatial or temporal dependence from
training data~\cite{dualCNN,dongDeepCNN,mehrabiPredictor}. These
are strong tracking methods, particularly when the assumed dynamics remain
matched to the physical link. Nevertheless, a tracker alone does not specify
the complete lifecycle of a persistent wireless state. The receiver still
needs to decide what to store, how uncertainty affects the next recovery,
whether delayed future pilots may refine an earlier state, and whether a new
candidate is sufficiently reliable to overwrite memory. A persistent ET must
make these operations explicit without assuming that one temporal model is
correct for every propagation event.

We address this lifecycle through a common correction-based inverse problem.
New pilots first measure the mismatch between the physical channel and the ET
prediction. Stored uncertainty and support confidence then determine how
strongly each coefficient is allowed to depart from memory. For immediate CSI,
the resulting convex problem is solved by uncertainty-aware ET FISTA
(UET-FISTA). When a short reporting delay is acceptable, the same model is
applied to an anchored window and solved by ET-ADMM, allowing later pilot
blocks to refine earlier channel states. These are not two unrelated
estimators: they are zero-delay and fixed-lag operating modes of the same
persistent-state formulation.

Updating memory is treated separately from fitting a candidate. The fitting
pilots cannot provide an impartial test of the estimate they produced, so a
small held-out pilot set evaluates whether the candidate generalizes to
unseen measurements. Depending on this evidence, the receiver incrementally
updates the ET, performs a complete static refresh, or holds the previous
state. The resulting loop has a direct wireless interpretation. The ET
predicts before sounding, corrects itself from sparse pilots, validates what
it has learned, stores only an accepted state, and then answers a CSI or beam
query. New pilots at the next epoch close the loop. Thus, \emph{persistence}
does not mean that an old channel is always preferred; it means that channel
memory is repeatedly confronted with physical evidence.

The proposed formulation also exposes when persistence should help. If only a
small part of the channel changes, recovering the mismatch can require fewer
pilots than reconstructing the entire channel. If the support changes
substantially or the stored uncertainty is poorly calibrated, that advantage
shrinks and may disappear. The validation action is therefore part of the
estimator rather than an implementation detail. This viewpoint leads to a
more limited but testable claim than universal tracking superiority: an ET
provides a reusable interface for CSI memory, uncertainty, delayed refinement,
and safe fallback, while the best underlying temporal model may still depend
on the propagation regime.

This paper makes the following contributions.
\begin{itemize}
\item We define a queryable ET state and distinguish its per-epoch
synchronization role from CKM mapping and a broader wireless digital twin.
The derived observation model shows that new pilots recover the total mismatch
between the current channel and stored prediction, including accumulated state
error rather than only ideal physical innovation.
\item We formulate one anchored-window inverse problem whose specializations
give static acquisition, zero-delay ET correction, and fixed-lag smoothing.
\item We derive low-latency UET-FISTA and fixed-lag ET-ADMM as two solvers of
that objective family. ET-OMP is retained as a low-complexity benchmark rather
than introduced as a separate estimation model.
\item We calibrate update/refresh/hold decisions with unseen pilots, derive
finite validation and contraction conditions, and give conditional online
and fixed-lag Bayesian bounds. Controlled 200-trajectory and independent
QuaDRiGa tests isolate smooth tracking, abrupt recovery, off-grid leakage,
and effective matched-beam rate under equal pilot budgets.
\end{itemize}


\section{Positioning CKM, Digital Twin, and ET}
\subsection{Three Objects With Different Scopes}
Let $\mathbf z$ denote a location or an environment descriptor. A CKM can be
abstracted as a learned mapping from $\mathbf z$ to a channel mean and
covariance,
\begin{equation}
\mathcal M_{\rm CKM}:\mathbf z
\longmapsto
\left(\mathbf m_{\rm ch}(\mathbf z),
\mathbf V_{\rm ch}(\mathbf z)\right),
\label{eq:ckm-map}
\end{equation}
where $\mathbf m_{\rm ch}(\mathbf z)$ and $\mathbf V_{\rm ch}(\mathbf z)$
represent the location-conditioned channel description and its uncertainty.
The output may instead be path loss, a beam index, another channel statistic,
or a channel representation~\cite{zengCKMTutorial}. The map may be updated over
time, but a per-sounding recursive synchronization law is not required by the
CKM definition.

A wireless digital twin is a broader virtual state. For illustration, we write
\begin{equation}
\mathcal D_t=
\left(
\mathcal G_t,\mathcal N_t,\mathcal R_t,\mathcal C_t,
\mathcal T_t,\ldots
\right),
\label{eq:dt-container}
\end{equation}
where $\mathcal G_t$, $\mathcal N_t$, $\mathcal R_t$, and $\mathcal C_t$ can represent geometry, network entities, resources, and control state, respectively. The electromagnetic state $\mathcal T_t$ considered in this paper can be one component of $\mathcal D_t$; it is not claimed to reproduce the complete network.

The proposed ET is the recursive state estimator
\begin{equation}
\mathcal T_t
=\mathcal U_{\rm ET}
\left(
\mathcal T_{t-1},\mathbf y_t,\mathbf\Phi_t
\right),
\qquad
\mathbf q_t=\mathcal Q_{\rm ET}(\mathcal T_t),
\label{eq:et-update-query}
\end{equation}
where $\mathcal U_{\rm ET}$ synchronizes stored electromagnetic knowledge with current pilots and $\mathcal Q_{\rm ET}$ returns a CSI, beamforming, or uncertainty query. If a CKM is available, it enters through initialization,
\begin{equation}
\widehat{\mathbf x}_{0|-1}
=\mathbf m_{\rm ch}(\mathbf z_0),\qquad
\mathbf V_{0|-1}
=\mathbf V_{\rm ch}(\mathbf z_0),
\label{eq:ckm-initialization}
\end{equation}
or through a slower prior refresh. After initialization, pilot evidence drives the fast ET recursion.

\subsection{Static Sparse Channel Estimation}
For an $N$-dimensional sparse channel with $K$ resolvable paths, a static
compressed-sensing estimator solves a new recovery problem from the current
pilot block. Orthogonal matching pursuit (OMP) offers a greedy solution, while
$\ell_1$ regularization admits convex solvers such as FISTA
~\cite{troppOMP,beckFISTA}. These estimators remain important rather than
obsolete: they provide the complete-refresh candidate whenever the stored ET
state is inaccurate, and they establish how much gain comes from persistence
rather than from the current pilots alone.

\subsection{Dynamic Recovery and Learned Temporal Priors}
Kalman-filtered compressed sensing estimates persistent coefficients on a
current support and searches the filtering residual for additions. Dynamic
message passing propagates support and amplitude beliefs, while temporally
correlated SBL learns hierarchical source variances
~\cite{vaswaniKFCS,zinielDCS,zhangTSBL}. Modified-CS and online
sparsity-aware RLS instead reuse partial support or recursively weighted
observations~\cite{vaswaniModifiedCS,angelosanteRLS}. Learned estimators can
also infer spatial and temporal dependence from
data~\cite{dualCNN,dongDeepCNN,mehrabiPredictor}. These are strong
tracking methods and are used as such in the experiments. The distinction is
not that an ET is the only possible temporal estimator. Rather, the ET exposes
an externally queryable state and an evidence-based choice among incremental
correction, complete refresh, and hold when a selected temporal model becomes
unreliable.

\begin{table*}[t]
\centering
\caption{CKM, Digital Twin, and ET}
\label{tab:positioning}
\renewcommand{\arraystretch}{1.04}
\scriptsize
\begin{tabular}{
>{\raggedright\arraybackslash}p{0.13\textwidth}
>{\raggedright\arraybackslash}p{0.23\textwidth}
>{\raggedright\arraybackslash}p{0.24\textwidth}
>{\raggedright\arraybackslash}p{0.23\textwidth}}
\toprule
Aspect & CKM & Wireless digital twin & Proposed ET in this paper\\
\midrule
Primary object &
Descriptor-to-propagation mapping &
Virtual network/asset representation &
Persistent CSI, uncertainty, confidence, and age\\
Typical input &
Location, geometry, images, or historical measurements &
Telemetry, geometry, traffic, sensing, and control data &
Previous state and current pilots $(\mathbf y_t,\mathbf\Phi_t)$\\
Typical output &
Channel statistic, radio-map value, beam, or prior &
Monitoring, simulation, prediction, or control state &
Updated CSI, reliability, and a channel-dependent query\\
Required time recursion &
Optional; it may be offline or slowly updated &
At the timescale of the twinned system &
Every epoch: predict--correct--validate--store\\
Role here &
Initializer or slow propagation prior &
Umbrella architecture that may contain the ET &
Signal-processing state and update problem\\
\bottomrule
\end{tabular}
\end{table*}

\subsection{Boundary of the Claimed Contribution}
Table~\ref{tab:positioning} contrasts the proposed ET with a CKM and a wireless
digital twin along their primary objects, inputs, outputs, and update
timescales. It also prevents two overclaims. First, this paper does not
rename a CKM: neither spatial interpolation nor complete radio-map
construction is required by the proposed inverse problem. Second, it does not
claim a complete network digital twin: traffic, mobility, computing, and
network-control states remain outside $\mathcal T_t$. The contribution is the
persistent electromagnetic synchronization law, its online and fixed-lag
solvers, and its reliability mechanism. CKM construction and
full network-twin orchestration remain compatible external modules.

\section{Persistent ET Channel Model}
\subsection{One Update in Plain Terms}
At the beginning of epoch $t$, the receiver already has a predicted channel
$\widehat{\mathbf x}_{t|t-1}$. It does not estimate the complete
$\mathbf x_t$ from zero. Instead, it asks for the correction
\begin{equation}
\mathbf u_t
=\underbrace{\mathbf x_t}_{\text{current physical channel}}
-\underbrace{\widehat{\mathbf x}_{t|t-1}}_{\text{ET prediction}}.
\label{eq:plain-correction}
\end{equation}
The pilots reveal this correction through a residual. After estimating
$\mathbf u_t$, the receiver forms one candidate channel, checks the resulting candidate
on pilots not used for estimation, and decides what to store. The complete
epoch can therefore be summarized as
\begin{align}
\text{predict:}\quad&
\widehat{\mathbf x}_{t|t-1}
=\mathcal P_t(\mathcal T_{t-1}),
\label{eq:step-predict}\\
\text{measure:}\quad&
\mathbf r_t
=\mathbf y_t-\mathbf\Phi_t\widehat{\mathbf x}_{t|t-1},
\label{eq:step-measure}\\
\text{correct:}\quad&
\widehat{\mathbf u}_t
=\mathcal A_t(\mathbf r_t,\mathbf\Phi_t,\mathcal T_{t-1}),
\label{eq:step-correct}\\
\text{store:}\quad&
\mathcal T_t
=\mathcal G_t(\text{update},\text{refresh},\text{hold}),
\label{eq:step-store}\\
\text{query:}\quad&
\mathbf q_t
=\mathcal Q_t(\mathcal T_t).
\label{eq:step-query}
\end{align}
Sections~IV--VI specify $\mathcal A_t$, while Section~VII specifies
$\mathcal G_t$. The remaining subsections define each physical variable used
by these five steps.

\begin{figure*}[t]
\centering
\includegraphics[width=0.98\textwidth]{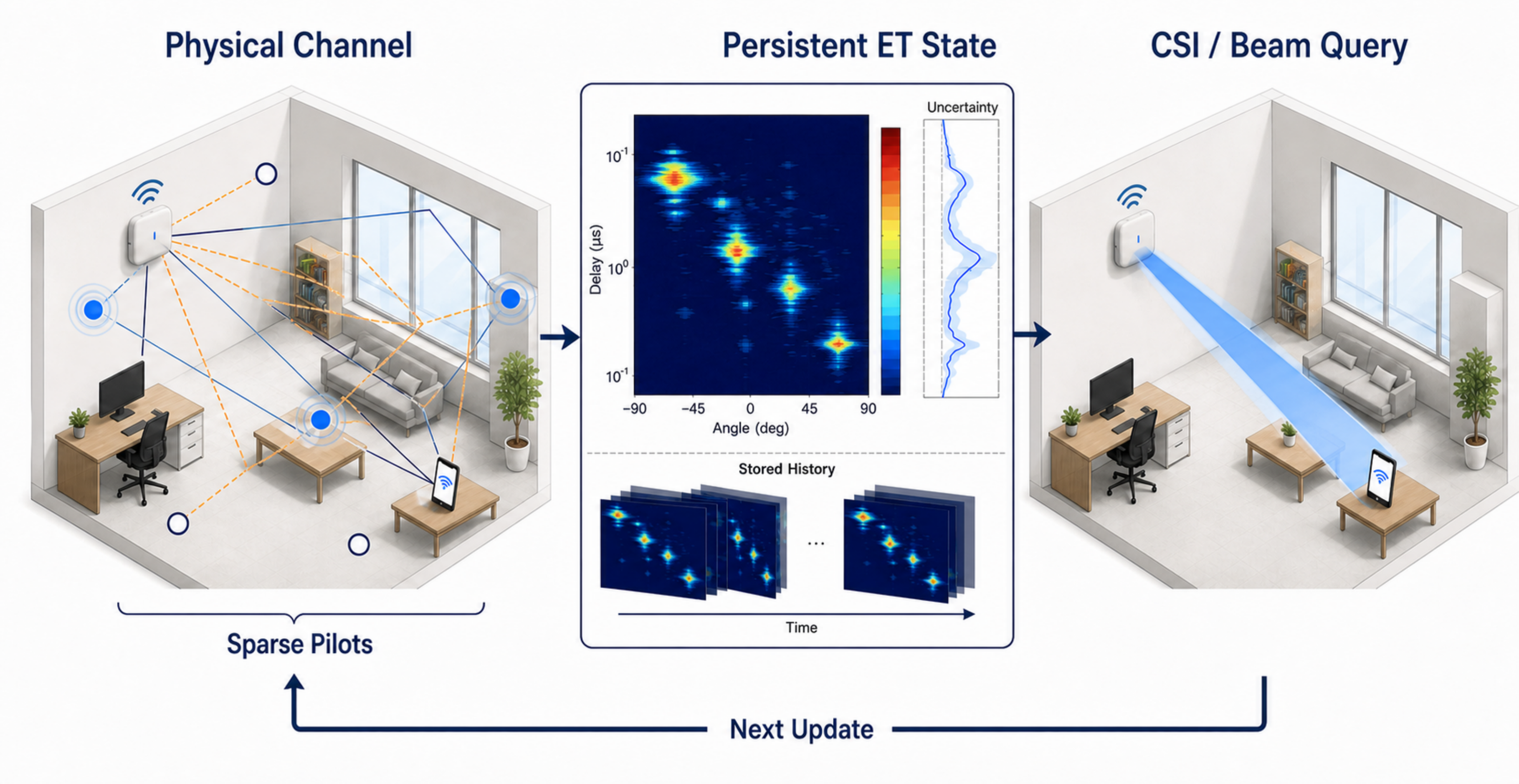}
\caption{Physical ET update loop.}
\label{fig:physical-model}
\end{figure*}

Fig.~\ref{fig:physical-model} connects the abstract operators in
\eqref{eq:step-predict}--\eqref{eq:step-query} to a wireless deployment. A
selected location contributes one or more rows to $\mathbf\Phi_t$ and the
corresponding entries of $\mathbf y_t$; unselected locations incur no current
pilot cost. The central panel visualizes angle--delay power, uncertainty, and
stored history, not a requirement to interpolate a complete spatial CKM. The
right-hand query illustrates one use of the updated CSI. Selection in the
figure only distinguishes measured from candidate probes; an optimized pilot
selector is outside the present claim and can use the stored uncertainty in a
future active-sounding extension.

\subsection{Physical Angle--Delay State}
Consider an angle--delay dictionary with $N=N_\theta N_\tau$ bins. At epoch $t$, the channel is
\begin{equation}
\mathbf x_t\in\C^N,\qquad \|\mathbf x_t\|_0=K_t,
\label{eq:physical-state}
\end{equation}
where each nonzero coefficient represents a resolvable propagation path. The physical innovation is
\begin{equation}
\boldsymbol\delta_t=\mathbf x_t-\mathbf x_{t-1}.
\label{eq:physical-innovation}
\end{equation}
Local gain changes, path births, path deaths, and blockage often make $\boldsymbol\delta_t$ sparse or compressible over a sufficiently short update interval.

\subsection{Twin Prediction and Correction}
Before receiving epoch-$t$ pilots, the ET predicts
\begin{equation}
\widehat{\mathbf x}_{t|t-1}
=\mathbf F_t\widehat{\mathbf x}_{t-1|t-1},
\label{eq:prediction}
\end{equation}
where $\mathbf F_t=\mathbf I$ gives a persistence predictor and a learned or
Doppler-aware transition can be substituted without changing the measurement
interface. After the fitting pilots arrive, a low-dimensional specialization
can align a common gain and phase before coefficientwise correction. Let
\begin{equation}
\mathbf z_t=\mathbf\Phi_{{\rm tr},t}\widehat{\mathbf x}_{t-1|t-1},
\qquad
\widetilde\alpha_t=
\frac{\mathbf z_t^H\mathbf y_{{\rm tr},t}}
{\|\mathbf z_t\|_2^2+\epsilon_\alpha},
\label{eq:scalar-predictor}
\end{equation}
and set
\begin{equation}
\widehat\alpha_t=
\left[|\widetilde\alpha_t|\right]_{\alpha_{\min}}^{\alpha_{\max}}
\exp\!\big(j\angle\widetilde\alpha_t\big),
\qquad
\mathbf F_t=\widehat\alpha_t\mathbf I,
\label{eq:bounded-scalar-predictor}
\end{equation}
where $[a]_\ell^u=\min\{u,\max\{\ell,a\}\}$ and
$\epsilon_\alpha>0$ stabilizes a weak projected state. Only fitting pilots
enter \eqref{eq:scalar-predictor}; held-out pilots remain unseen by both this
predictor and the correction solver. This one-complex-scalar fit removes a
common Doppler phase while leaving path-dependent changes to
$\mathbf u_t$. Define the previous estimation error
\begin{equation}
\mathbf e_{t-1}=
\widehat{\mathbf x}_{t-1|t-1}-\mathbf x_{t-1}.
\end{equation}
For $\mathbf F_t=\mathbf I$, the correction required by the receiver is
\begin{equation}
\mathbf u_t
=\mathbf x_t-\widehat{\mathbf x}_{t|t-1}
=\boldsymbol\delta_t-\mathbf e_{t-1}.
\label{eq:correction}
\end{equation}
Equation~\eqref{eq:correction} explains both the gain and the risk of memory. A sparse physical innovation does not guarantee a sparse correction if the stored error is dense. Conversely, a calibrated twin converts full-channel estimation into a lower-dimensional correction problem.

\begin{proposition}[Correction support]
\label{prop:correction-support}
If $\|\boldsymbol\delta_t\|_0\leq k_{\Delta,t}$ and
$\|\mathbf e_{t-1}\|_0\leq k_{e,t-1}$, then
\begin{equation}
\|\mathbf u_t\|_0
\leq k_{\Delta,t}+k_{e,t-1}.
\label{eq:correction-support}
\end{equation}
The result follows directly from
$\supp(\mathbf u_t)\subseteq
\supp(\boldsymbol\delta_t)\cup\supp(\mathbf e_{t-1})$.
Thus the ET reduces pilot demand only when the right-hand side of
\eqref{eq:correction-support} remains below the effective sparsity of the complete channel.
\end{proposition}

\subsection{Sparse Pilot Observation}
After pilot combining, the receiver collects $M_t<N$ observations modeled as
\begin{equation}
\mathbf y_t=\mathbf\Phi_t\mathbf x_t+\mathbf n_t,
\qquad
\mathbf n_t\sim\mathcal{CN}(\mathbf 0,\sigma_n^2\mathbf I),
\label{eq:pilot}
\end{equation}
where $\mathbf\Phi_t\in\C^{M_t\times N}$ is known. Subtracting the ET prediction gives the exact residual model
\begin{equation}
\mathbf r_t
=\mathbf y_t-\mathbf\Phi_t\widehat{\mathbf x}_{t|t-1}
=\mathbf\Phi_t\mathbf u_t+\mathbf n_t.
\label{eq:residual}
\end{equation}
Unlike treating $\mathbf e_{t-1}$ only as additional noise, \eqref{eq:residual} recognizes that current pilots can correct a sparse component of stored-state error.

\subsection{Stored Uncertainty and Support Confidence}
The ET state is
\begin{equation}
\mathcal T_t=
\left(
\widehat{\mathbf x}_{t|t},
\mathbf v_{t|t},
\mathbf p_{t|t},
a_t
\right),
\label{eq:twin-state}
\end{equation}
where $\mathbf v_{t|t}\in\R_+^N$ contains marginal error variances,
$\mathbf p_{t|t}\in[0,1]^N$ contains support confidences, and $a_t$ is the number of epochs since a complete refresh. A diagonal prediction is
\begin{equation}
\mathbf v_{t|t-1}
=\rho_t^2\mathbf v_{t-1|t-1}
+q_0\mathbf 1+q_1\mathbf p_{t-1|t-1}.
\label{eq:variance-prediction}
\end{equation}
The floor $q_0$ permits new paths, while $q_1$ assigns additional uncertainty to previously active paths whose gains may change.

After an accepted correction, a local Gaussian approximation to the
data-consistency term gives
\begin{equation}
\mathbf V_{t|t}^{\rm post}
\approx
\left[
(\mathbf V_{t|t-1})^{-1}
+\frac{\mathbf\Phi_t^H\mathbf\Phi_t}{\sigma_n^2}
+\mu_t\mathbf I
\right]^{-1},
\label{eq:posterior-covariance}
\end{equation}
where $\mathbf V_{t|t-1}=\diag(\mathbf v_{t|t-1})$. Maintaining the full
matrix is unnecessary for the proposed coefficientwise weighting. A diagonal
implementation uses
\begin{equation}
v_{t|t,i}
=
\left[
\frac{1}{v_{t|t-1,i}}
+\frac{\|\mathbf\Phi_t(:,i)\|_2^2}{\sigma_n^2}
+\mu_t
\right]^{-1}.
\label{eq:diagonal-posterior}
\end{equation}
Thus, a coefficient becomes more certain when the current pilots carry more
energy along its dictionary atom.

Support confidence is updated separately because a small posterior variance
does not imply that a coefficient is active. Define the soft activity score
\begin{equation}
s_{t,i}
=
\frac{1}{
1+\exp[-(|\widehat x_{t|t,i}|-\tau_t)/\zeta_t]},
\label{eq:activity-score}
\end{equation}
where $\tau_t$ is an activity threshold and $\zeta_t$ controls its softness.
The stored confidence is
\begin{equation}
p_{t|t,i}
=(1-\eta_p)p_{t|t-1,i}+\eta_p s_{t,i},
\qquad 0<\eta_p\leq1.
\label{eq:confidence-update}
\end{equation}
Equations~\eqref{eq:diagonal-posterior} and
\eqref{eq:confidence-update} serve different purposes: the former measures
how accurately a coefficient is known, while the latter measures how likely
it is to belong to the channel support.

\begin{remark}[Separate uncertainty and activity]
\label{rem:uncertainty-activity}
The ET must store both uncertainty and activity. A path can
be confidently absent, confidently present, or present with an uncertain
gain; one scalar ``channel confidence'' cannot distinguish these cases.
Table~\ref{tab:symbols} summarizes the stored variables and their role in the
next prediction, correction, and validation cycle.
\end{remark}

\begin{table}[t]
\centering
\caption{ET State Variables}
\label{tab:symbols}
\begin{tabular}{cl}
\toprule
Symbol & Meaning\\
\midrule
$\mathbf x_t$ & True channel at epoch $t$\\
$\boldsymbol\delta_t$ & Physical change $\mathbf x_t-\mathbf x_{t-1}$\\
$\widehat{\mathbf x}_{t|t-1}$ & Prediction before current pilots\\
$\mathbf u_t$ & Correction $\mathbf x_t-\widehat{\mathbf x}_{t|t-1}$\\
$\mathcal T_t$ & State stored after the update\\
\bottomrule
\end{tabular}
\end{table}

\section{A Unified Persistent-State Estimation Problem}
\subsection{Single-Epoch ET-Conditioned Prior}
The exact residual model~\eqref{eq:residual} gives the likelihood
\begin{equation}
p(\mathbf r_t|\mathbf u_t)
\propto
\exp\left(
-\frac{
\|\mathbf r_t-\mathbf\Phi_t\mathbf u_t\|_2^2
}{2\sigma_n^2}
\right).
\label{eq:correction-likelihood}
\end{equation}
To convert the ET state into a prior, assign coefficient $i$ the predicted
change scale
\begin{equation}
b_{t,i}
=
\frac{1}{c_t}
\sqrt{
\epsilon+\alpha p_{t|t-1,i}
+(1-\alpha)v_{t|t-1,i}/\bar v_t
},
\label{eq:change-scale}
\end{equation}
where $\bar v_t=N^{-1}\sum_i v_{t|t-1,i}$ and $c_t$ normalizes the median
inverse scale to one. A previously active or uncertain coefficient receives a
larger $b_{t,i}$ because a correction there is more plausible. The inverse
scale is
\begin{equation}
w_{t,i}
=\frac{1}{b_{t,i}}
=\frac{c_t}{
\sqrt{\epsilon+\alpha p_{t|t-1,i}
+(1-\alpha)v_{t|t-1,i}/\bar v_t}},
\label{eq:weights}
\end{equation}
and the independent elastic-net prior is
\begin{equation}
p(\mathbf u_t|\mathcal T_{t-1})
\propto
\prod_{i=1}^{N}
\exp\left(
-\lambda_t w_{t,i}|u_{t,i}|
-\frac{\mu_t}{2}|u_{t,i}|^2
\right).
\label{eq:correction-prior}
\end{equation}
The weighted Laplace factor promotes sparse corrections, while the weak
Gaussian factor gives the elastic-net stabilization used for nearly collinear
pilot atoms~\cite{candesReweighted,zouElastic}. The objective is strongly
convex when $\mu_t>0$.

Taking the negative logarithm of
$p(\mathbf r_t|\mathbf u_t)p(\mathbf u_t|\mathcal T_{t-1})$ and discarding
constants gives the MAP problem
\begin{equation}
\begin{aligned}
\widehat{\mathbf u}_t
=\arg\min_{\mathbf u\in\C^N}\quad&
\frac{1}{2\sigma_n^2}
\|\mathbf r_t-\mathbf\Phi_t\mathbf u\|_2^2\\
&+\lambda_t\|\mathbf W_t\mathbf u\|_1
+\frac{\mu_t}{2}\|\mathbf u\|_2^2,
\end{aligned}
\label{eq:online-problem}
\end{equation}
where $\mathbf W_t=\diag(\mathbf w_t)$ and
$\mathbf w_t=[w_{t,1},\ldots,w_{t,N}]^T$. Thus,
problem~\eqref{eq:online-problem} follows from the Gaussian pilot likelihood
and the ET-conditioned prior rather than being introduced only as a convenient
regularizer.

\begin{remark}[A low weight permits change]
\label{rem:weight-meaning}
A large $p_{t|t-1,i}$ or $v_{t|t-1,i}$ reduces
$w_{t,i}$ and lowers the threshold on $u_{t,i}$. This does not mean that an
uncertain coefficient is trusted more; it means that current pilots are
allowed to correct it more aggressively. A confidently absent coefficient is
changed only when the residual provides strong evidence.
\end{remark}

\subsection{Anchored Window Formulation}
To place online correction, static recovery, and fixed-lag smoothing under one
objective, let $a=t-L+1$ and collect
\[
\mathbf X_t=
[\mathbf x_a,\ldots,\mathbf x_t]\in\C^{N\times L}.
\]
The first transition is anchored to the ET prediction preceding the window:
\begin{align}
\mathbf d_a
&=\mathbf x_a-\widehat{\mathbf x}_{a|a-1},\label{eq:first-transition}\\
\mathbf d_\tau
&=\mathbf x_\tau-\mathbf F_\tau\mathbf x_{\tau-1},
\quad \tau=a+1,\ldots,t. \label{eq:internal-transition}
\end{align}
Define the affine transition operator
$\mathcal D_{F,a}(\mathbf X_t)=[\mathbf d_a,\ldots,\mathbf d_t]$.
For diagonal $\mathbf W_\tau\succeq\mathbf0$ constructed from the predicted
ET state and $\mathbf R_\tau\succeq\mathbf0$, consider
\begin{equation}
\begin{aligned}
\widehat{\mathbf X}_t
=\arg\min_{\mathbf X_t}\quad&
\sum_{\tau=a}^{t}
\frac{1}{2\sigma_n^2}
\|\mathbf y_\tau-\mathbf\Phi_\tau\mathbf x_\tau\|_2^2\\
&+\lambda_s\|\mathbf X_t\|_1
+\lambda_d\sum_{\tau=a}^{t}
\|\mathbf W_\tau\mathbf d_\tau\|_1\\
&+\frac{1}{2}\sum_{\tau=a}^{t}
\|\mathbf d_\tau\|_{\mathbf R_\tau}^2,
\end{aligned}
\label{eq:persistent-problem}
\end{equation}
where $\|\mathbf z\|_{\mathbf R}^2=\mathbf z^H\mathbf R\mathbf z$.
The four terms impose pilot consistency, optional sparsity of the complete
channel, sparse ET-conditioned corrections, and quadratic stabilization of
those corrections. The first transition connects every window to the stored
state; the remaining transitions form an ET-conditioned complex fused-lasso
model~\cite{tibshiraniFused}. Problem~\eqref{eq:persistent-problem} is convex
for fixed nonnegative weights and positive-semidefinite
$\{\mathbf R_\tau\}$.

\begin{proposition}[Exact operating-mode reductions]
\label{prop:master-reductions}
Problem~\eqref{eq:persistent-problem} contains the three estimators used in
this paper.
\begin{enumerate}
\item For $L=1$, $\lambda_s=0$, $\lambda_d=\lambda_t$,
$\mathbf W_t$ from \eqref{eq:weights}, and
$\mathbf R_t=\mu_t\mathbf I$, substituting
$\mathbf x_t=\widehat{\mathbf x}_{t|t-1}+\mathbf u_t$ gives
\eqref{eq:online-problem} exactly.
\item For $L=1$, $\mathbf W_t=\mathbf0$, $\mathbf R_t=\mathbf0$, and
$\lambda_s>0$, it becomes static sparse channel recovery.
\item For $L>1$, nonzero transition weights couple the window and yield the
fixed-lag ET smoother solved by ET-ADMM.
\end{enumerate}
\end{proposition}

To verify the first statement, note that the data residual is
$\mathbf y_t-\mathbf\Phi_t\mathbf x_t
=\mathbf r_t-\mathbf\Phi_t\mathbf u_t$, while the last two terms become
$\lambda_t\|\mathbf W_t\mathbf u_t\|_1+
\mu_t\|\mathbf u_t\|_2^2/2$. Hence, no objective is changed when moving from
the online to the window notation.

\begin{remark}[Latency and evidence]
\label{rem:latency-evidence}
UET-FISTA and ET-ADMM solve operating modes of
\eqref{eq:persistent-problem} under different latency requirements.
UET-FISTA returns $\widehat{\mathbf x}_{t|t}$ immediately. Fixed-lag smoothing
waits for up to $L-1$ additional pilot blocks and can revise earlier states.
Increasing $L$ supplies more evidence, but it also raises complexity and can
make the sparse-difference model less accurate if the propagation environment
changes substantially within the window.
\end{remark}

\section{Online ET Solvers}
\subsection{Convex UET-FISTA}
We write the online objective as $F_t(\mathbf u)=f_t(\mathbf u)+g_t(\mathbf u)$,
where
\[
f_t(\mathbf u)=
\frac{\|\mathbf r_t-\mathbf\Phi_t\mathbf u\|_2^2}{2\sigma_n^2}
+\frac{\mu_t}{2}\|\mathbf u\|_2^2
\]
is smooth and
$g_t(\mathbf u)=\lambda_t\sum_iw_{t,i}|u_i|$ is nonsmooth but separable.
Using the complex gradient convention for a real-valued objective,
\begin{equation}
\nabla f_t(\mathbf u)
=
\frac{\mathbf\Phi_t^H
(\mathbf\Phi_t\mathbf u-\mathbf r_t)}{\sigma_n^2}
+\mu_t\mathbf u.
\label{eq:smooth-gradient}
\end{equation}
For any $\mathbf u$ and $\mathbf z$, the quadratic upper bound
\begin{equation}
\begin{aligned}
f_t(\mathbf u)\leq\;&f_t(\mathbf z)
+\Re\{\nabla f_t(\mathbf z)^H(\mathbf u-\mathbf z)\}\\
&+\frac{L_t}{2}\|\mathbf u-\mathbf z\|_2^2
\end{aligned}
\label{eq:smooth-majorizer}
\end{equation}
holds with
\begin{equation}
L_t=\frac{\|\mathbf\Phi_t\|_2^2}{\sigma_n^2}+\mu_t.
\label{eq:lipschitz}
\end{equation}
Minimizing the sum of \eqref{eq:smooth-majorizer} and $g_t$ is equivalent to
the proximal problem
\begin{equation}
\arg\min_{\mathbf u}\;
\frac{1}{2}
\left\|
\mathbf u-
\left(\mathbf z-\frac{\nabla f_t(\mathbf z)}{L_t}\right)
\right\|_2^2
+\frac{\lambda_t}{L_t}
\sum_iw_{t,i}|u_i|.
\label{eq:proximal-subproblem}
\end{equation}
It separates across coefficients. For a complex scalar $z_i$, phase
alignment is optimal, and minimizing over its magnitude yields
\begin{equation}
\soft_{\boldsymbol\gamma}(\mathbf z)_i
=\max\left(0,1-\frac{\gamma_i}{|z_i|}\right)z_i.
\label{eq:complex-soft}
\end{equation}
Substituting $\gamma_i=\lambda_tw_{t,i}/L_t$ into
\eqref{eq:complex-soft} gives the UET-FISTA proximal-gradient update
\begin{equation}
\mathbf d^{(q+1)}
=\soft_{\lambda_t\mathbf w_t/L_t}
\left[
\mathbf z^{(q)}-\frac{\nabla f_t(\mathbf z^{(q)})}{L_t}
\right],
\label{eq:uet-fista}
\end{equation}
followed by
\begin{equation}
\begin{aligned}
\beta_{q+1}&=\frac{1+\sqrt{1+4\beta_q^2}}{2},\\
\mathbf z^{(q+1)}
&=\mathbf d^{(q+1)}
+\frac{\beta_q-1}{\beta_{q+1}}
(\mathbf d^{(q+1)}-\mathbf d^{(q)}).
\end{aligned}
\end{equation}
After convergence, define a reproducible correction support by
\begin{equation}
\widehat{\mathcal S}_t
=
\left\{
i:\; |d_i|\geq\tau_s\max_j|d_j|
\right\},
\qquad
|\widehat{\mathcal S}_t|\leq k_{\max},
\label{eq:support-rule}
\end{equation}
where the $k_{\max}$ largest coefficients are retained if the threshold
selects more than the correction budget. Shrinkage bias is removed by
\begin{equation}
\widehat{\mathbf u}_t(\widehat{\mathcal S}_t)
=\mathbf\Phi_t(:,\widehat{\mathcal S}_t)^\dagger\mathbf r_t.
\label{eq:debias}
\end{equation}
The channel update is
\begin{equation}
\widehat{\mathbf x}^{\rm upd}_{t|t}
=\widehat{\mathbf x}_{t|t-1}+\widehat{\mathbf u}_t.
\label{eq:online-update}
\end{equation}

\begin{remark}[Where the ET enters UET-FISTA]
\label{rem:uet-reduction}
UET-FISTA does not alter the measured residual or invent
additional observations. The ET affects only the coefficientwise proximal
thresholds. If all $w_{t,i}=1$, the method reduces to ordinary elastic-net
FISTA on the correction; if the prediction is zero, the correction equals the
complete channel and the formulation reduces to static sparse estimation.
\end{remark}

\begin{algorithm}[t]
\caption{Online UET-FISTA Update}
\label{alg:uet}
\begin{algorithmic}[1]
\STATE Predict $\widehat{\mathbf x}_{t|t-1}$ and $\mathbf v_{t|t-1}$.
\STATE Form $\mathbf r_t$ using \eqref{eq:residual} and $\mathbf w_t$ using \eqref{eq:weights}.
\STATE Initialize $\mathbf d^{(0)}=\mathbf z^{(0)}=\mathbf 0$ and $\beta_0=1$.
\REPEAT
\STATE Update $\mathbf d^{(q+1)}$ using \eqref{eq:uet-fista}.
\STATE Update momentum $\beta_{q+1}$ and extrapolated point $\mathbf z^{(q+1)}$.
\UNTIL{the relative correction change is below $\varepsilon_{\rm F}$}
\STATE Debias on $\widehat{\mathcal S}_t$ using \eqref{eq:debias}.
\STATE Form candidate $\widehat{\mathbf x}^{\rm upd}_{t|t}$ using \eqref{eq:online-update}.
\end{algorithmic}
\end{algorithm}

\begin{proposition}[Online convergence]
\label{prop:online-convergence}
Because $f_t$ is convex with an $L_t$-Lipschitz gradient and
$\lambda_t\|\mathbf W_t\mathbf u\|_1$ is proper, closed, and convex, the objective values generated by UET-FISTA satisfy
\begin{equation}
F_t(\mathbf d^{(q)})-F_t(\mathbf u_t^\star)
\leq
\frac{2L_t\|\mathbf d^{(0)}-\mathbf u_t^\star\|_2^2}{(q+1)^2},
\label{eq:fista-rate}
\end{equation}
before the optional debiasing step~\cite{beckFISTA}. The weights affect the proximal threshold but do not destroy convexity.
\end{proposition}

\subsection{Low-Complexity ET-OMP}
ET-OMP applies OMP to \eqref{eq:residual}. At each iteration it selects the column with the largest residual correlation, refits least squares on the selected support, and stops after a correction budget or a residual test. Its dominant complexity is
$\mathcal O(k_uM_tN)$, compared with
$\mathcal O(I_{\rm F}M_tN)$ for $I_{\rm F}$ UET-FISTA iterations. ET-OMP is retained as a low-complexity operating point rather than presented as a different physical problem.

\section{Fixed-Lag ET-ADMM}
Let $\mathbf x=\vect(\mathbf X_t)$ and define
\begin{align}
\boldsymbol\Psi_t
&=\blkdiag(\mathbf\Phi_a,\ldots,\mathbf\Phi_t),\\
\mathbf D_{F,t}\mathbf x-\mathbf c_t
&=\vect(\mathcal D_{F,a}(\mathbf X_t)),\label{eq:affine-transition}\\
\mathbf c_t
&=[\widehat{\mathbf x}_{a|a-1}^T,\mathbf0^T,\ldots,\mathbf0^T]^T,
\end{align}
where $\mathbf D_{F,t}$ is block lower bidiagonal, with $\mathbf I$ on its
diagonal and $-\mathbf F_\tau$ on its first block subdiagonal. Also let
$\mathbf R_t^{(L)}=\blkdiag(\mathbf R_a,\ldots,\mathbf R_t)$ and
$\mathbf W_t^{(L)}=\blkdiag(\mathbf W_a,\ldots,\mathbf W_t)$. The smooth
part of \eqref{eq:persistent-problem} is
\begin{equation}
h_t(\mathbf x)=
\frac{\|\mathbf y_t^{(L)}-\boldsymbol\Psi_t\mathbf x\|_2^2}
{2\sigma_n^2}
+\frac{1}{2}\|\mathbf D_{F,t}\mathbf x-\mathbf c_t\|_
{\mathbf R_t^{(L)}}^2,
\label{eq:admm-smooth-part}
\end{equation}
where $\mathbf y_t^{(L)}=[\mathbf y_a^T,\ldots,\mathbf y_t^T]^T$.
Introduce
\begin{equation}
\mathbf z=\mathbf x,\qquad
\mathbf u=\mathbf D_{F,t}\mathbf x-\mathbf c_t.
\label{eq:admm-splits}
\end{equation}
Problem~\eqref{eq:persistent-problem} becomes
\begin{equation}
\begin{aligned}
\min_{\mathbf x,\mathbf z,\mathbf u}\quad&
h_t(\mathbf x)+\lambda_s\|\mathbf z\|_1
+\lambda_d\|\mathbf W_t^{(L)}\mathbf u\|_1\\
\text{s.t.}\quad&
\mathbf x-\mathbf z=\mathbf0,\qquad
\mathbf D_{F,t}\mathbf x-\mathbf c_t-\mathbf u=\mathbf0.
\end{aligned}
\label{eq:admm-constrained}
\end{equation}
With scaled dual variables $\mathbf y_z,\mathbf q$ and penalty $\rho>0$,
the complete scaled augmented Lagrangian is
\begin{equation}
\begin{aligned}
\mathcal L_\rho
=\;&h_t(\mathbf x)+\lambda_s\|\mathbf z\|_1
+\lambda_d\|\mathbf W_t^{(L)}\mathbf u\|_1\\
&+\frac{\rho}{2}\|\mathbf x-\mathbf z+\mathbf y_z\|_2^2
-\frac{\rho}{2}\|\mathbf y_z\|_2^2\\
&+\frac{\rho}{2}
\|\mathbf D_{F,t}\mathbf x-\mathbf c_t-\mathbf u+\mathbf q\|_2^2
-\frac{\rho}{2}\|\mathbf q\|_2^2.
\end{aligned}
\label{eq:scaled-augmented-lagrangian}
\end{equation}
This expression makes the two nonsmooth structures explicit:
$\mathbf z$ carries within-epoch channel sparsity and $\mathbf u$ carries
sparse transition corrections. Differentiating
\eqref{eq:scaled-augmented-lagrangian} with respect to $\mathbf x$ gives
\begin{equation}
\mathbf H_t\mathbf x^{(q+1)}=\mathbf b_t^{(q)},
\label{eq:block-system}
\end{equation}
where
\begin{equation}
\mathbf H_t
=\frac{\boldsymbol\Psi_t^H\boldsymbol\Psi_t}{\sigma_n^2}
+\mathbf D_{F,t}^H\mathbf R_t^{(L)}\mathbf D_{F,t}
+\rho\mathbf I+\rho\mathbf D_{F,t}^H\mathbf D_{F,t}
\label{eq:admm-hessian}
\end{equation}
and
\begin{equation}
\begin{aligned}
\mathbf b_t^{(q)}
=\;&\frac{\boldsymbol\Psi_t^H\mathbf y_t^{(L)}}{\sigma_n^2}
+\mathbf D_{F,t}^H\mathbf R_t^{(L)}\mathbf c_t
+\rho(\mathbf z^{(q)}-\mathbf y_z^{(q)})\\
&+\rho\mathbf D_{F,t}^H
(\mathbf u^{(q)}-\mathbf q^{(q)}+\mathbf c_t).
\end{aligned}
\label{eq:admm-rhs}
\end{equation}
Because transitions couple only adjacent epochs, $\mathbf H_t$ is block
tridiagonal. It is handled by matrix-free conjugate gradients or, for a short
window, one reusable Cholesky factorization. Epoch-dependent noise variances
only replace the first Gram term by its block-weighted counterpart.

The remaining subproblems are elementwise:
\begin{align}
\mathbf z^{(q+1)}
&=\soft_{\lambda_s/\rho}
(\mathbf x^{(q+1)}+\mathbf y_z^{(q)}),
\label{eq:admm-z}\\
\mathbf u^{(q+1)}
&=\soft_{\lambda_d\mathbf w_t^{(L)}/\rho}
(\mathbf D_{F,t}\mathbf x^{(q+1)}-\mathbf c_t+\mathbf q^{(q)}),
\label{eq:admm-u}
\end{align}
where $\mathbf w_t^{(L)}=\diag(\mathbf W_t^{(L)})$. The dual updates are
\begin{align}
\mathbf y_z^{(q+1)}
&=\mathbf y_z^{(q)}+\mathbf x^{(q+1)}-\mathbf z^{(q+1)},\\
\mathbf q^{(q+1)}
&=\mathbf q^{(q)}+\mathbf D_{F,t}\mathbf x^{(q+1)}
-\mathbf c_t-\mathbf u^{(q+1)}.
\end{align}
The two constraint residuals can be monitored jointly through
\begin{equation}
r_{\rm p}^{(q)}
=\left(
\|\mathbf x^{(q)}-\mathbf z^{(q)}\|_2^2+
\|\mathbf D_{F,t}\mathbf x^{(q)}-\mathbf c_t-\mathbf u^{(q)}\|_2^2
\right)^{1/2},
\label{eq:admm-primal-residual}
\end{equation}
while the scaled dual residual is
\begin{equation}
\begin{aligned}
\big(r_{\rm d}^{(q)}\big)^2
=\;&\rho^2\|\mathbf z^{(q)}-\mathbf z^{(q-1)}\|_2^2\\
&+\rho^2\|\mathbf D_{F,t}^H
(\mathbf u^{(q)}-\mathbf u^{(q-1)})\|_2^2.
\end{aligned}
\label{eq:admm-dual-residual}
\end{equation}
Iterations stop when \eqref{eq:admm-primal-residual} and
\eqref{eq:admm-dual-residual} satisfy the standard absolute and relative
tolerances~\cite{boydADMM}. The $\rho\mathbf I$ term in
\eqref{eq:admm-hessian} makes the state-update Hessian positive definite for
every $\rho>0$; hence each $\mathbf x$ subproblem has a unique solution. The
latest column is current CSI, while earlier columns are corrected fixed-lag
states.

\begin{remark}[Causal meaning of fixed-lag smoothing]
\label{rem:fixed-lag-causality}
ET-ADMM gains accuracy by letting a future pilot block
correct a recent past state through the temporal-difference term. It does not
use future information without accounting for delay: the maximum delay is
explicitly $L-1$ epochs. The online mode corresponds to zero additional delay,
whereas the fixed-lag mode trades delay and computation for joint evidence.
When the validation gate selects a complete refresh, the current window is
closed and the refreshed state anchors a new segment. Consequently, ET-ADMM
does not propagate a pre-change transition model across an event already
rejected by current evidence.
\end{remark}

\begin{algorithm}[t]
\caption{Fixed-Lag ET-ADMM}
\label{alg:admm}
\begin{algorithmic}[1]
\STATE Build the latest $L$-epoch window and anchor it to $\mathcal T_{t-L}$.
\STATE Warm-start $\mathbf x$ from online UET-FISTA states.
\REPEAT
\STATE Solve the block system \eqref{eq:block-system} for $\mathbf x^{(q+1)}$.
\STATE Apply \eqref{eq:admm-z} and \eqref{eq:admm-u}.
\STATE Update scaled dual variables $\mathbf y_z$ and $\mathbf q$.
\UNTIL{primal and dual residuals satisfy ADMM tolerances}
\STATE Return the latest state and store corrected lagged states.
\end{algorithmic}
\end{algorithm}

\begin{proposition}[ET-ADMM convergence]
\label{prop:admm-convergence}
For fixed nonnegative $\mathbf W_t^{(L)}$, positive-semidefinite
$\mathbf R_t^{(L)}$, $\rho>0$, and $\sigma_n^2>0$,
problem~\eqref{eq:persistent-problem} is proper, closed, and convex. If its
solution set is nonempty, the primal residuals of Algorithm~\ref{alg:admm}
converge to zero and its objective converges to the optimum~\cite{boydADMM}.
The fixed-lag mode therefore solves the same persistent-state objective family
as the online mode, rather than introducing a different physical problem.
\end{proposition}

\section{Reliable Persistent Updating}
\label{sec:reliable-updating}
\subsection{Update, Refresh, or Hold}
Partition the current pilots into a training subset and a validation subset
$(\mathbf y_{v,t},\mathbf\Phi_{v,t})$ containing $M_v$ measurements. The
validation pilots are excluded from UET-FISTA, ET-ADMM, and the static refresh
estimator. Construct three candidates:
\begin{align}
\widehat{\mathbf x}^{\rm upd}_{t|t}
&=\widehat{\mathbf x}_{t|t-1}+\widehat{\mathbf u}_t,\\
\widehat{\mathbf x}^{\rm ref}_{t|t}
&=\text{a static estimate from current training pilots},\\
\widehat{\mathbf x}^{\rm hold}_{t|t}
&=\widehat{\mathbf x}_{t|t-1}.
\end{align}
Their validation errors are
\begin{equation}
\epsilon_{c,t}
=\|\mathbf y_{v,t}-\mathbf\Phi_{v,t}
\widehat{\mathbf x}^{c}_{t|t}\|_2^2,
\quad c\in\{\mathrm{upd},\mathrm{ref},\mathrm{hold}\}.
\label{eq:three-errors}
\end{equation}
Under Gaussian validation noise,
$\epsilon_{c,t}/\sigma_n^2$ is proportional to the negative
log-likelihood of candidate $c$. Define the normalized score
\begin{equation}
\bar\epsilon_{c,t}
=\frac{\epsilon_{c,t}}{M_v\sigma_n^2}
\label{eq:normalized-validation}
\end{equation}
and let $h$ denote the hold candidate. Conditional on the training data, all
three candidates are fixed and independent of the validation noise. For
$c\in\{\mathrm{upd},\mathrm{ref}\}$, define
\begin{equation}
D_{c,h}
=\bar\epsilon_{c,t}-\bar\epsilon_{h,t}.
\label{eq:validation-difference}
\end{equation}
Let
$\mathbf a_{c,h}
=\mathbf\Phi_{v,t}
(\widehat{\mathbf x}^{c}_{t|t}-\widehat{\mathbf x}^{h}_{t|t})$.
Expanding the two squared errors cancels the common noise energy and gives
\begin{equation}
D_{c,h}|\mathcal F_{\rm tr}
\sim\mathcal N\!\left(
\mu_{c,h},
\frac{2\|\mathbf a_{c,h}\|_2^2}
{M_v^2\sigma_n^2}
\right),
\label{eq:validation-difference-law}
\end{equation}
where $\mathcal F_{\rm tr}$ is the training-data sigma-field and
$\mu_{c,h}$ is the normalized difference between their noise-free validation
losses. For a desired family-wise false-acceptance probability
$\alpha_v$, use the Bonferroni margin
\begin{equation}
m_{c,h}
=
\frac{\sqrt{2}\,z_{1-\alpha_v/2}
\|\mathbf a_{c,h}\|_2}
{M_v\sigma_n},
\label{eq:calibrated-margin}
\end{equation}
where $z_p$ is the standard-normal quantile. Candidate $c$ is eligible only
if
\begin{equation}
\bar\epsilon_{c,t}+m_{c,h}<\bar\epsilon_{h,t}.
\label{eq:eligible-candidate}
\end{equation}
Let $\mathcal C_t$ contain the update and refresh candidates satisfying
\eqref{eq:eligible-candidate}. The state manager chooses the eligible
candidate with smaller validation score; if neither is eligible, it holds:
\begin{equation}
d_t=
\begin{cases}
\displaystyle\arg\min_{c\in\mathcal C_t}
\bar\epsilon_{c,t},&
\mathcal C_t\neq\varnothing,\\
\mathrm{hold},&\text{otherwise}.
\end{cases}
\label{eq:gate-decision}
\end{equation}

\begin{proposition}[False-acceptance control]
\label{prop:false-acceptance}
Suppose the validation noise is proper complex Gaussian and independent of
candidate construction. If an adaptive candidate $c$ is no better than hold
in noise-free validation loss, then
\[
\Pr\!\left[
D_{c,h}<-m_{c,h}\mid\mathcal F_{\rm tr}
\right]\leq\alpha_v/2.
\]
Consequently, the probability that either update or refresh is accepted even
though it is no better than hold is at most $\alpha_v$.
\end{proposition}
This follows directly from \eqref{eq:validation-difference-law} and a union
bound. If
$\bar a_{c,h}^2=M_v^{-1}\|\mathbf a_{c,h}\|_2^2$, requiring a maximum
normalized margin $\tau_v$ gives the finite-sample condition
\begin{equation}
M_v\geq
\frac{2z_{1-\alpha_v/2}^2\bar a_{c,h}^2}
{\tau_v^2\sigma_n^2}.
\label{eq:validation-sample-requirement}
\end{equation}
Thus, the number of synchronization pilots is tied to a declared risk and
resolution, rather than selected only as a plotting parameter.
The stored channel is then
\begin{equation}
\widehat{\mathbf x}_{t|t}
=
\begin{cases}
\widehat{\mathbf x}^{\rm upd}_{t|t},&d_t=\mathrm{upd},\\
\widehat{\mathbf x}^{\rm ref}_{t|t},&d_t=\mathrm{ref},\\
\widehat{\mathbf x}_{t|t-1},&d_t=\mathrm{hold}.
\end{cases}
\label{eq:selected-state}
\end{equation}
For an update, $\mathbf v_{t|t}$ follows
\eqref{eq:diagonal-posterior}; for a refresh, it is reset from the static
estimator's local curvature; for hold, it remains
$\mathbf v_{t|t-1}$. State age evolves as
\begin{equation}
a_t=
\begin{cases}
0,&d_t=\mathrm{ref},\\
a_{t-1}+1,&d_t\in\{\mathrm{upd},\mathrm{hold}\}.
\end{cases}
\label{eq:state-age}
\end{equation}
Support confidence is recomputed from \eqref{eq:activity-score} after update
or refresh and is mildly decayed after hold.

\begin{remark}[Validation without ground truth]
\label{rem:validation-overhead}
The gate never observes the true channel. It compares only
how three independently constructed candidates predict unseen pilots. Its
$M_v$ measurements must be counted as synchronization overhead; otherwise,
the reliability gain would be obtained through an unfair pilot advantage.
After the decision is fixed, the selected non-hold model may be refitted on
the union of training and validation pilots. This refit cannot alter $d_t$ and
does not reuse validation data to choose its own model.
\end{remark}

\subsection{Long-Term Error Interpretation}
For an accepted incremental update,
\begin{equation}
\begin{aligned}
\widehat{\mathbf x}^{\rm upd}_{t|t}-\mathbf x_t
&=
\widehat{\mathbf x}_{t|t-1}+\widehat{\mathbf u}_t-\mathbf x_t\\
&=
\widehat{\mathbf u}_t-\mathbf u_t.
\end{aligned}
\label{eq:error-cancellation}
\end{equation}
Thus, previous state error is not automatically added again after every
epoch. It affects the next estimate by making $\mathbf u_t$ less sparse or by
miscalibrating its weights. If the current correction is recovered accurately,
\eqref{eq:error-cancellation} corrects both physical innovation and a
compressible part of the previous error.

Let $E_t=\|\widehat{\mathbf x}_{t|t}-\mathbf x_t\|_2$ and let
$\sigma_s(\mathbf u)_1$ denote the $\ell_1$ error of the best $s$-term
approximation of $\mathbf u$. Suppose the accepted update operator obeys
\begin{equation}
E_t\leq\kappa_tE_{t-1}
+c_0\frac{\sigma_s(\mathbf u_t)_1}{\sqrt{s}}
+c_1\|\mathbf n_t\|_2,
\label{eq:error-recursion}
\end{equation}
where $\sigma_s(\mathbf u)_1$ is the best $s$-term approximation error. Define
\[
b_t=c_0\frac{\sigma_s(\mathbf u_t)_1}{\sqrt{s}}
+c_1\|\mathbf n_t\|_2.
\]
Repeated substitution of \eqref{eq:error-recursion} gives the nonstationary
finite-time relation
\begin{equation}
E_t\leq
\left(\prod_{j=1}^{t}\kappa_j\right)E_0+
\sum_{i=1}^{t}
\left(\prod_{j=i+1}^{t}\kappa_j\right)b_i.
\label{eq:nonstationary-error-recursion}
\end{equation}
Thus a small present error alone is not sufficient for persistence: the
sequence of accepted contraction factors must remain controlled. If
$\kappa_j\leq\bar\kappa<1$ and $b_i\leq b$, the products reduce to a geometric
series,
\begin{equation}
E_t\leq
\bar\kappa^tE_0+
b\sum_{\ell=0}^{t-1}\bar\kappa^\ell
=\bar\kappa^tE_0+\frac{1-\bar\kappa^t}{1-\bar\kappa}b.
\label{eq:bounded-error}
\end{equation}
Equation~\eqref{eq:bounded-error} is a sufficient stability condition, not an unconditional guarantee. It clarifies the role of the validation gate: updates that do not empirically contract prediction error should be replaced by refresh or hold decisions.

\begin{corollary}[Gate risk and expected contraction]
\label{cor:gate-contraction}
Let $\mathcal E_t$ denote the event that update or refresh is falsely accepted.
Suppose the selected decision has conditional expected factor
$\kappa_g<1$ on $\mathcal E_t^c$, factor
$\kappa_b>\kappa_g$ on $\mathcal E_t$, and an additive term with conditional
mean at most $b$. Under Proposition~\ref{prop:false-acceptance},
\begin{equation}
\mathbb E[E_t|\mathcal F_{t-1}]
\leq
\underbrace{\big[(1-\alpha_v)\kappa_g+
\alpha_v\kappa_b\big]}_{\bar\kappa_v}
E_{t-1}+b.
\label{eq:gate-contraction}
\end{equation}
A sufficient expected-contraction condition is
\begin{equation}
\alpha_v<
\frac{1-\kappa_g}{\kappa_b-\kappa_g}.
\label{eq:gate-risk-condition}
\end{equation}
\end{corollary}
This condition is deliberately sufficient: it exposes how statistical gate
risk enters persistence, but it requires refresh, hold, and correctly accepted
updates collectively to satisfy the stated good-event factor; it does not
claim that every channel trajectory obeys one fixed pair
$(\kappa_g,\kappa_b)$.

\begin{remark}[Persistence can repair stored error]
\label{rem:error-repair}
Persistence does not mean blindly accumulating estimates.
Equation~\eqref{eq:error-cancellation} shows that each pilot block can repair
the stored state. Error accumulates only when the correction is not
recoverable or a poor candidate repeatedly passes validation; these are the
two failure modes exposed by the stored-state and gate experiments.
\end{remark}

\subsection{Complexity}
Table~\ref{tab:complexity} compares all implemented practical estimators,
rather than only the proposed modes. UET-FISTA has the same leading order as
static FISTA and remains in the same millisecond-scale range as KF-CS in this
implementation. Sequential SBL repeats an $M\times M$ factorization and is
slower, whereas direct ET-ADMM pays substantially more for a joint four-epoch
solution. For large $LN$, ET-ADMM can replace the direct factorization by
matrix-free preconditioned conjugate gradients (PCG), with dominant cost
$\mathcal O(I_{\rm A}I_{\rm CG}LMN)$.

\begin{table*}[t]
\centering
\caption{Per-Update Complexity and Solver-Only Runtime}
\label{tab:complexity}
\small
\setlength{\tabcolsep}{8pt}
\begin{tabular}{lccc}
\toprule
Method & Dominant cost & Median time (ms) & 95th percentile (ms)\\
\midrule
Static FISTA & $\mathcal O(I_{\rm F}MN)$ & 1.14 & 1.60\\
KF-CS & $\mathcal O(MN+M^3+M^2K)$ & 0.55 & 0.90\\
Sequential SBL & $\mathcal O(I_{\rm S}(M^3+M^2N))$ & 2.66 & 4.90\\
ET-OMP & $\mathcal O(k_uMN+Mk_u^2+k_u^3)$ & 0.09 & 0.14\\
UET-FISTA & $\mathcal O(I_{\rm F}MN)$ & 1.08 & 1.21\\
ET-ADMM (direct) &
$\mathcal O((LN)^3+I_{\rm A}(LN)^2)$ & 32.55 & 45.79\\
\bottomrule
\end{tabular}
\parbox{0.94\textwidth}{\footnotesize The benchmark uses the paper's
$N=128$, 34 total pilots, 26 fitting pilots, and $L=4$ window on an Intel Core
Ultra 7 265K workstation with MATLAB R2024b and one worker. Values average
100 calls after four warm-up calls per method. They include only the listed
solver, not channel generation, held-out validation, state management, or
plotting.}
\end{table*}

\section{ET-Aware Performance Bounds}
\subsection{Conditional Bayesian CRLB for One Update}
The classical deterministic CRLB is not directly informative for the full
$N$-dimensional correction when $M_t<N$. We therefore condition on a local
correction support
$\mathcal S_t=\supp(\mathbf u_t)$ with
$s_t=|\mathcal S_t|\leq M_t$ and derive an optimistic benchmark for the
remaining coefficient-estimation error. Conditioned on
$(\mathcal S_t,\mathcal T_{t-1})$, we use the local Gaussian uncertainty model
\begin{equation}
\mathbf u_{t,\mathcal S_t}
\sim\mathcal{CN}(\mathbf0,\mathbf V_{u,t}),\qquad
\mathbf r_t
=\mathbf\Phi_{t,\mathcal S_t}\mathbf u_{t,\mathcal S_t}+\mathbf n_t,
\label{eq:bound-local-model}
\end{equation}
where $\mathbf V_{u,t}\succ\mathbf0$ is obtained from the predicted ET
uncertainty restricted to $\mathcal S_t$. The negative log joint density,
up to a constant and an immaterial common scale, is
\begin{equation}
\mathcal L_t(\mathbf u)
=
\frac{\|\mathbf r_t-\mathbf\Phi_{t,\mathcal S_t}\mathbf u\|_2^2}
{\sigma_n^2}
+\mathbf u^H\mathbf V_{u,t}^{-1}\mathbf u.
\label{eq:bound-joint-density}
\end{equation}
Its data and prior information matrices are, respectively,
\begin{equation}
\mathbf J_{d,t}
=\frac{\mathbf\Phi_{t,\mathcal S_t}^H
\mathbf\Phi_{t,\mathcal S_t}}{\sigma_n^2},
\qquad
\mathbf J_{p,t}=\mathbf V_{u,t}^{-1}.
\label{eq:data-prior-information}
\end{equation}

\begin{proposition}[Conditional ET Bayesian CRLB]
\label{prop:online-bcrlb}
For the local model in \eqref{eq:bound-local-model}, every estimator satisfying
the Bayesian regularity conditions obeys~\cite{kayEstimation,vanTreesBounds}
\begin{equation}
\mathbb E\!\left[
(\widehat{\mathbf u}_{t,\mathcal S_t}-\mathbf u_{t,\mathcal S_t})
(\widehat{\mathbf u}_{t,\mathcal S_t}-\mathbf u_{t,\mathcal S_t})^H
\mid\mathcal S_t,\mathcal T_{t-1}
\right]
\succeq
\mathbf B_t,
\label{eq:online-bcrlb}
\end{equation}
where
\begin{equation}
\mathbf B_t
=
\left(
\mathbf V_{u,t}^{-1}
+\frac{\mathbf\Phi_{t,\mathcal S_t}^H
\mathbf\Phi_{t,\mathcal S_t}}{\sigma_n^2}
\right)^{-1}.
\label{eq:bcrlb-matrix}
\end{equation}
\end{proposition}

The result follows because the Bayesian information is the sum
$\mathbf J_{p,t}+\mathbf J_{d,t}$. Moreover, the posterior mean
\begin{equation}
\widehat{\mathbf u}_{t,\mathcal S_t}^{\rm post}
=\mathbf B_t
\frac{\mathbf\Phi_{t,\mathcal S_t}^H\mathbf r_t}{\sigma_n^2}
\label{eq:oracle-posterior}
\end{equation}
attains \eqref{eq:online-bcrlb} for the linear Gaussian model. Since the channel
update error equals the correction error by \eqref{eq:error-cancellation}, the
normalized lower bound reported in the simulations is
\begin{equation}
\beta_t^{\rm ET}
=
\frac{\tr(\mathbf B_t)}
{\mathbb E[\|\mathbf x_t\|_2^2]}.
\label{eq:normalized-bcrlb}
\end{equation}
Removing ET information, i.e.,
$\mathbf V_{u,t}^{-1}\rightarrow\mathbf0$, gives the data-only
oracle-support CRLB
\begin{equation}
\mathbf B_t^{\rm data}
=\sigma_n^2
\left(
\mathbf\Phi_{t,\mathcal S_t}^H\mathbf\Phi_{t,\mathcal S_t}
\right)^{-1}.
\label{eq:data-only-crlb}
\end{equation}

\begin{remark}[What the bound does and does not measure]
\label{rem:bound-scope}
The ET-BCRLB is conditioned on the true local support and uses a Gaussian
surrogate for the stored uncertainty. It quantifies the information available
after the support ambiguity has been removed; it is not a claim that
UET-FISTA knows $\mathcal S_t$, nor does it describe the bias introduced by
soft thresholding. The gap from UET-FISTA to \eqref{eq:normalized-bcrlb}
therefore contains support-selection error, prior mismatch, shrinkage, and
finite-iteration error.
\end{remark}

\subsection{Fixed-Lag Smoothing Bound}
To isolate the information benefit of fixed-lag processing, replace the
weighted Laplace transition prior in \eqref{eq:persistent-problem} by the
local Gaussian model
\begin{equation}
\mathbf d_a\sim\mathcal{CN}(\mathbf0,\mathbf V_a),\qquad
\mathbf d_\tau\sim\mathcal{CN}(\mathbf0,\mathbf Q_d),
\quad \tau=a+1,\ldots,t.
\label{eq:gaussian-window-prior}
\end{equation}
Let
$\overline{\mathbf Q}_L=\blkdiag(\mathbf V_a,\mathbf Q_d,\ldots,\mathbf Q_d)$
and retain $\boldsymbol\Psi_t$ and $\mathbf D_{F,t}$ from
\eqref{eq:affine-transition}. The window information matrix is
\begin{equation}
\mathbf J_t^{(L)}
=
\frac{\boldsymbol\Psi_t^H\boldsymbol\Psi_t}{\sigma_n^2}
+\mathbf D_{F,t}^H\overline{\mathbf Q}_L^{-1}\mathbf D_{F,t}.
\label{eq:window-information}
\end{equation}

\begin{proposition}[Fixed-lag Bayesian CRLB]
\label{prop:lag-bcrlb}
Under \eqref{eq:gaussian-window-prior}, the covariance of any regular
window estimator satisfies~\cite{tichavskyPCRLB}
\begin{equation}
\operatorname{Cov}\!\left[
\vect(\widehat{\mathbf X}_t-\mathbf X_t)
\right]
\succeq
\left(\mathbf J_t^{(L)}\right)^{-1}.
\label{eq:window-bcrlb}
\end{equation}
If $\mathbf E_\ell$ selects state $\ell$ in the window, its marginal bound is
\begin{equation}
\mathbf B_{\ell,t}^{(L)}
=
\mathbf E_\ell
\left(\mathbf J_t^{(L)}\right)^{-1}
\mathbf E_\ell^H.
\label{eq:marginal-window-bound}
\end{equation}
\end{proposition}

\begin{corollary}[Why smoothing helps a recent past state]
\label{cor:smoothing-information}
For a fixed state in the window, conditioning on an additional future pilot
block cannot increase its Bayesian minimum mean-square error. Because the
linear Gaussian posterior covariance attains
\eqref{eq:marginal-window-bound}, the corresponding smoothed BCRLB cannot
increase either. The improvement is largest when $\mathbf Q_d$ strongly
couples adjacent epochs and saturates when process uncertainty or pilot noise
weakens that coupling.
\end{corollary}

\begin{remark}[Bound--algorithm connection]
\label{rem:bound-algorithm}
The quadratic terms of \eqref{eq:window-information} have the same
block-tridiagonal structure as the ET-ADMM state update in
\eqref{eq:admm-hessian}. The bound uses Gaussian local dynamics to quantify
information, whereas ET-ADMM keeps the sparse $\ell_1$ temporal model to remain
robust to path births, deaths, and abrupt blockage.
\end{remark}

\section{Numerical Results}
\subsection{Protocols and Reproducibility}
We use a controlled $N_\theta\times N_\tau=16\times8$ angle--delay grid with
$K=10$ initial paths. The first protocol isolates one update and averages 500
independent channel and pilot realizations. Unless varied, four coefficients
change, the pilot ratio is 20\%, and the pilot signal-to-noise ratio (SNR) is
10 dB. Minimum-norm LS, static OMP, static FISTA, and a stale prediction are
practical references; oracle innovation LS is used only to expose the loss
caused by unknown correction support.

The second protocol evaluates persistent operation over 50 epochs and 200
independent trajectories. Each ordinary epoch contains three changed
coefficients, epochs 15 and 39 replace two paths, and epoch 28 attenuates half
of the active paths and applies a phase discontinuity. Tuning and test seeds
are disjoint. Every practical method receives the same $M_{\rm tot}=34$
pilots at 10 dB pilot SNR. Static FISTA, Kalman-filtered compressed sensing
(KF-CS), and sequential SBL use all 34 pilots for estimation. Each managed ET
candidate uses 26 fitting pilots and eight held-out validation pilots; after a
non-hold decision, the selected model is refitted using all 34 pilots. Thus
validation changes how pilots are used but does not increase the total
overhead. The ET starts from a 20 dB-accuracy state, and fixed-lag ET-ADMM uses
$L=4$, corresponding to a three-epoch reporting delay.

Besides normalized mean-square error (NMSE) and support F1 score, we report
the effective rate of a normalized matched beam designed from the estimate:
\begin{equation}
R_{{\rm eff},t}=
\left(1-\frac{M_{\rm tot}}{T_c}\right)
\log_2\!\left(
1+\rho_d
\frac{|\mathbf x_t^H\widehat{\mathbf x}_t|^2}
{\|\mathbf x_t\|_2^2\|\widehat{\mathbf x}_t\|_2^2}
\right),
\label{eq:effective-rate}
\end{equation}
where $T_c=256$ symbols and $\rho_d=10$ dB. The true channel is used only for
evaluation. All curves are arithmetic means rather than selected trials. The
configuration, raw trajectory arrays, per-epoch CSV files, and fixed random
seeds are exported by the formal simulation runner.

\subsection{Single-Update Checks}
\begin{figure}[t]
\centering
\includegraphics[width=0.8\linewidth]{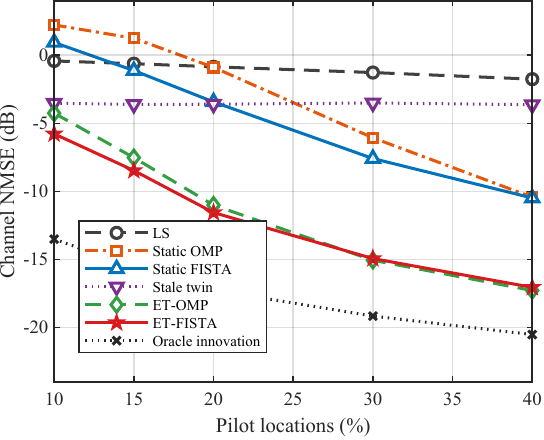}\\[-1mm]
\caption{NMSE versus pilot ratio.}
\label{fig:pilot-efficiency}
\end{figure}

\begin{figure}[t]
\centering
\includegraphics[width=0.8\linewidth]{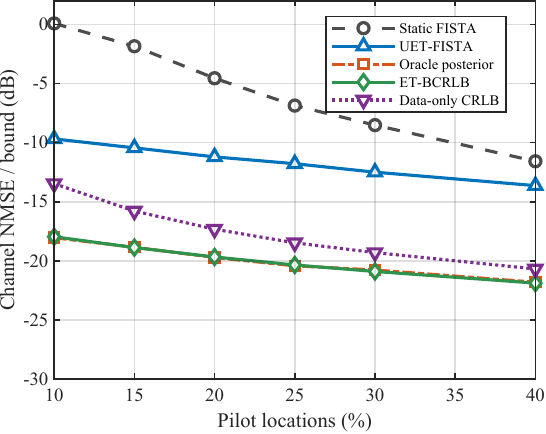}
\caption{Conditional ET-BCRLB verification.}
\label{fig:bound-verification}
\end{figure}

At 20\% pilots, Fig.~\ref{fig:pilot-efficiency} gives $-11.55$ dB NMSE for
ET-FISTA, compared with $-3.42$ dB for static FISTA, $-3.59$ dB for the stale
prediction, and $-11.00$ dB for ET-OMP. The stale result rules out an
uninteresting memory-only explanation: current pilots must estimate the
correction. As the stored state becomes inaccurate, the correction loses its
sparsity advantage and the result approaches static recovery. This
negative-transfer boundary motivates the refresh and hold actions in
Section~\ref{sec:reliable-updating}.

Fig.~\ref{fig:bound-verification} reports the bound experiment on the same $N=128$
grid, $K=10$, four changed
coefficients, a 15 dB predicted state, and 500 realizations per point.
At 20\% pilots, static FISTA, UET-FISTA, and the oracle posterior attain
$-4.55$, $-11.21$, and $-19.72$ dB, while the conditional ET-BCRLB is
$-19.68$ dB; the 0.04 dB agreement verifies \eqref{eq:bcrlb-matrix}. The
data-only bound is $-17.33$ dB, so calibrated stored information contributes
2.35 dB even after the local support is revealed. As emphasized in
Remark~\ref{rem:bound-scope}, this is an optimistic conditional reference,
not a lower bound for an estimator that must discover the support.

\subsection{Dynamic Tracking}
\begin{figure}[t]
\centering
\includegraphics[width=0.8\linewidth]{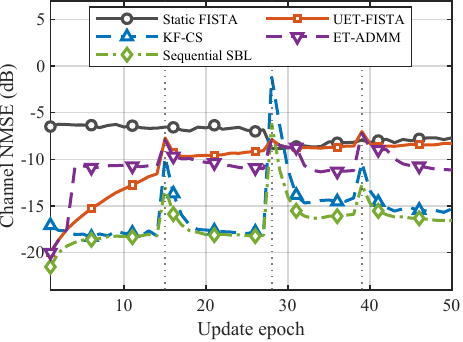}
\caption{Channel NMSE over 50 updates.}
\label{fig:tracking-nmse}
\end{figure}

\begin{figure}[t]
\centering
\includegraphics[width=0.8\linewidth]{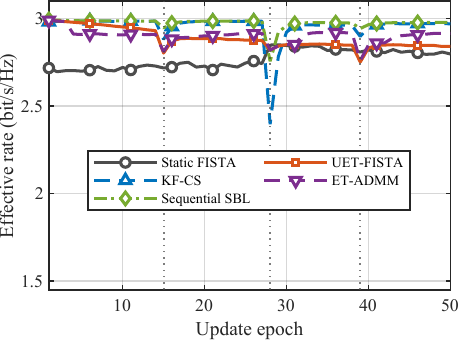}
\caption{Effective rate over 50 updates.}
\label{fig:tracking-rate}
\end{figure}

\begin{table*}[t]
\centering
\caption{Equal-Pilot Tracking Summary Over 200 Trajectories}
\label{tab:persistent-summary}
\small
\setlength{\tabcolsep}{7pt}
\begin{tabular}{lccccc}
\toprule
Method & Smooth NMSE (dB) & Epoch-28 NMSE (dB) & F1 &
Mean $R_{\rm eff}$ & Epoch-28 $R_{\rm eff}$\\
\midrule
Static FISTA & $-7.11$ & $-8.84$ & 0.706 & 2.766 & 2.847\\
KF-CS & $-16.36$ & $-1.15$ & 0.872 & 2.954 & 2.395\\
Sequential SBL & $-17.37$ & $-6.15$ & 0.931 & 2.972 & 2.755\\
Stale ET & $1.38$ & $5.04$ & 0.818 & 2.053 & 1.758\\
ET-OMP & $-10.30$ & $-8.15$ & 0.804 & 2.883 & 2.828\\
UET-FISTA & $-10.15$ & $-7.87$ & 0.815 & 2.883 & 2.819\\
Fixed-lag ET-ADMM (current) & $-10.91$ & $-7.87$ & 0.839 & 2.898 & 2.819\\
\bottomrule
\end{tabular}
\end{table*}

Figs.~\ref{fig:tracking-nmse} and~\ref{fig:tracking-rate}, together with
Table~\ref{tab:persistent-summary}, expose a
genuine model-matching tradeoff. Sequential SBL and KF-CS are the strongest
practical methods in the smooth regime, reaching $-17.37$ and $-16.36$ dB,
respectively. UET-FISTA reaches $-10.15$ dB and therefore does not dominate a
correctly matched temporal tracker. At the blockage epoch 28, however,
UET-FISTA attains $-7.87$ dB, compared with $-6.15$ dB for sequential SBL and
$-1.15$ dB for KF-CS. Each value is the ensemble average across 200
independent trajectories at that single epoch, not a temporal average over
several abrupt epochs. The corresponding effective rates are 2.819, 2.755, and
2.395 bit/s/Hz. The result supports a narrower claim: persistent correction
and validation sacrifice some smooth-model accuracy but reduce failure when a
stored temporal model becomes unreliable.

The two oracle diagnostics delimit the remaining gap. Oracle-support KF
attains $-18.45$ dB NMSE in smooth epochs but only $-4.46$ dB at blockage, whereas
oracle innovation recovery reaches $-18.52$ and $-12.01$ dB. Thus most smooth
loss is associated with practical support and variance learning, while the
abrupt gap also contains nonoracle change localization.

\subsection{Reliability and Smoothing}
The next experiment separates two consequences of persistence that a
time-averaged NMSE cannot reveal. First, it records whether unseen pilots
accept an update, request a complete refresh, or retain the stored state.
Second, it evaluates the same channel state online, at the current end of an
ADMM window, and after the declared three-epoch lag. A complete refresh
terminates the current smoothing segment, so a delayed estimate is not credited
with evidence propagated across an event rejected by the validation gate.
\begin{figure}[t]
\centering
\includegraphics[width=0.8\linewidth]{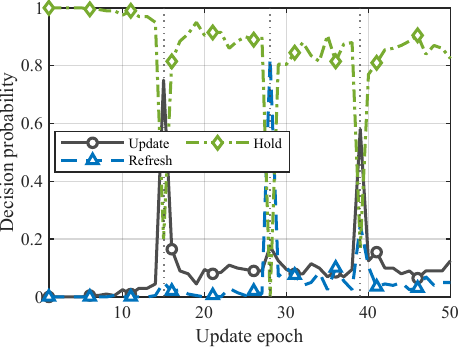}
\caption{UET update, refresh, and hold decisions.}
\label{fig:gate-probability}
\end{figure}

\begin{figure}[t]
\centering
\includegraphics[width=0.8\linewidth]{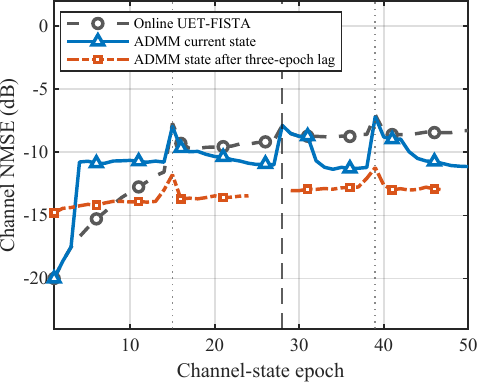}
\caption{Online and fixed-lag NMSE.}
\label{fig:fixed-lag-nmse}
\end{figure}

Figure~\ref{fig:gate-probability} shows that UET chooses update, refresh, and
hold with average probabilities 0.096, 0.049, and 0.855. At the blockage in
epoch 28, the refresh probability rises to 0.825; at the path events in epochs
15 and 39, update or refresh is selected on most trajectories. The high hold
probability in ordinary epochs is intentional: a fitted candidate replaces the
stored state only when unseen pilots provide statistically sufficient evidence.

Across 6,662 matched windows, Fig.~\ref{fig:fixed-lag-nmse} shows that online
UET, current-state ET-ADMM, and the
three-epoch-lagged ET-ADMM output attain $-10.28$, $-10.71$, and $-13.41$ dB NMSE,
respectively. Every ADMM window satisfies the declared primal and dual
tolerances. A direct block Cholesky update uses 26.1 iterations on average and
0.030 s per window on an Intel Core Ultra 7 265K workstation running MATLAB
R2024b with one worker; the 95th-percentile time is 0.034 s. These timings are
implementation- and hardware-dependent, but the matched-state NMSE confirms
the predicted accuracy--delay tradeoff independently of runtime.

\subsection{QuaDRiGa Validation}
We next replace the synthetic on-grid state by QuaDRiGa v2.8.1
\cite{jaeckelQuaDRiGa}, configured for the 3GPP UMi-NLOS model
\cite{3gpp38901}. A 16-element half-wavelength uniform linear array at
$(0,0,10)$ m serves a single-antenna user initialized at
$(20,-5,1.5)$ m. The carrier frequency and bandwidth are 28 GHz and 100 MHz.
The user moves at 1 m/s, pilot blocks arrive every 1 ms, and QuaDRiGa's
drifting mode preserves complex-path continuity over 50 epochs. Eight
subcarriers and a unitary $8\times16$ delay--angle transform again give
$N=128$. On average, 41.0\% of the transformed energy lies outside the ten
strongest bins, so this experiment does not satisfy exact grid sparsity.

Eight trajectories with seeds 12111--12118 select the physical-trace
regularization and drift scale; the 40 test trajectories use disjoint seeds
14111--14150. The total pilot, validation, pilot-SNR, and coherence budgets
remain 34, 8, 10 dB, and 256 symbols, respectively. Because off-grid leakage creates a
compressible rather than hard-sparse state, the physical protocol disables
support debiasing and permits 32 active correction coefficients. It also uses
\eqref{eq:bounded-scalar-predictor} with
$[\alpha_{\min},\alpha_{\max}]=[0.8,1.2]$. Every other decision is made from
the same pilots, and the true QuaDRiGa channel is used only for NMSE and
per-subcarrier maximum-ratio-transmission (MRT) rate evaluation.

\begin{figure}[t]
\centering
\includegraphics[width=0.8\linewidth]{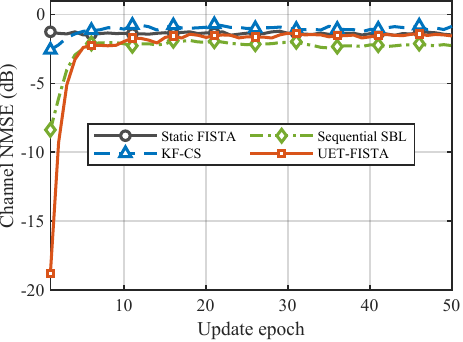}
\caption{QuaDRiGa channel NMSE.}
\label{fig:quadriga-nmse}
\end{figure}

\begin{figure}[t]
\centering
\includegraphics[width=0.8\linewidth]{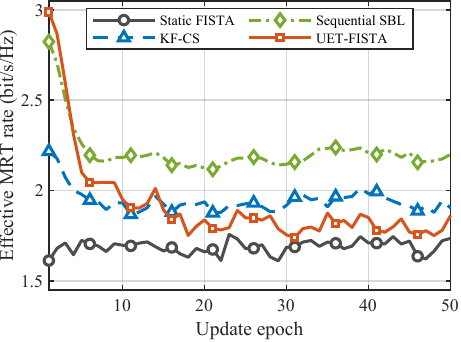}
\caption{QuaDRiGa effective MRT rate.}
\label{fig:quadriga-rate}
\end{figure}

Figure~\ref{fig:quadriga-nmse} gives $-1.883$ dB mean NMSE for UET-FISTA,
compared with $-1.371$, $-1.053$, and $-2.318$ dB for static FISTA, KF-CS,
and sequential SBL. Figure~\ref{fig:quadriga-rate} shows that the corresponding
effective rates are 1.919, 1.688, 1.943, and 2.215 bit/s/Hz. The physical
experiment therefore supports improvement over static recovery and robustness
to off-grid leakage, not universal dominance over a well-matched smooth
temporal model.

\section{Conclusion}
We have defined an electromagnetic twin as a persistent,
measurement-synchronized CSI state whose channel, uncertainty, support
confidence, and age are predicted, corrected, validated on unseen pilots, and
then exposed to a communication query. One anchored-window inverse problem
yields static refresh, online UET-FISTA, and fixed-lag ET-ADMM without
changing the underlying physical state. The results also delimit the claim. KF-CS and sequential SBL are more accurate
under correctly matched smooth dynamics, whereas correction-based UET-FISTA
is less sensitive to abrupt model failure. Fixed-lag processing supplies a
separate accuracy--delay mode, and calibrated decisions control whether new
evidence updates, refreshes, or preserves the state. Off-grid QuaDRiGa tests
show a gain over static recovery after pilot overhead, while SBL remains the
strongest smooth physical tracker.

\end{document}